# LOWER CRITICAL FIELD MEASUREMENT OF NbN MULTILAYER THIN FILM SUPERCONDUCTOR AT KEK

H. Ito[†], SOKENDAI, 305-0801 Tsukuba, Ibaraki, Japan
H. Hayano, T. Kubo, T. Saeki, R. Katayama, KEK, 305-0801 Tsukuba, Ibaraki, Japan
Y. Iwashita, H. Tongu, Kyoto ICR, Kyoto University, 611-0011 Uji, Kyoto, Japan
R. Ito, T. Nagata, ULVAC, Inc., 286-0225 Chiba, Japan
C. Z. Antoine, CEA, Irfu, F-91191 Gif-sur-Yvette, France

*Abstract*

The multilayer thin film structure of the superconductor has been proposed by A. Gurevich to enhance the maximum gradient of SRF cavities. The lower critical field $H_{c1}$ at which the vortex starts penetrating the superconducting material will be improved by coating Nb with thin film superconductor such as NbN. It is expected that the enhancement of $H_{c1}$ depends on the thickness of each layer. In order to determine the optimum thickness of each layer and to compare the measurement results with the theoretical prediction proposed by T. Kubo, we developed the $H_{c1}$ measurement system using the third harmonic response of the applied AC magnetic field at KEK. For the $H_{c1}$ measurement without the influence of the edge or the shape effects, the AC magnetic field can be applied locally by the solenoid coil of 5mm diameter in our measurement system. ULVAC made the NbN-SiO$_2$ multilayer thin film samples of various NbN thicknesses. In this report, the measurement result of the bulk Nb sample and NbN-SiO$_2$ multilayer thin film samples of different thickness of NbN layer will be discussed.

## INDRODUCTION

Superconductor-Insulator-Superconductor (S-I-S) thin film multilayer structure has been proposed by A. Gurevich to enhance the effective $H_{c1}$, and T. Kubo has proceeded with an advanced theoretical study to predict an optimum thickness of each layer which achieves the maximum effective $H_{c1}$ [1,2].

The effective $H_{c1}$ measurement for S-I-S structure sample must not be sensitive to a sample edge, at which the thickness of each layer cannot be guaranteed and a magnetic field is enhanced by the edge. Therefore, a new magnetic measurement system using third harmonic measurement method which can apply the magnetic field to the sample locally is necessary. In the case of the third harmonic measurement method, if a solenoid coil which applies the magnetic field to the sample is much smaller than the sample, the effective $H_{c1}$ of the sample can be measured directly without the edge effects.

In recent year, the effective $H_{c1}$ of the S-I-S structure sample has been measured by the third harmonic measurement method at CEA Saclay and Kyoto University, and then a possibility of enhancement of the effective $H_{c1}$ has been shown [3,4,5]. In addition, the third harmonic measurement system has been constructed at also KEK with a goal of measuring the S-I-S structure sample in a low-temperature region up to a liquid helium temperature [6,7]. Therefore, it is now at a stage where verification of theoretical prediction and search for optimum thickness of the S-I-S structure can be performed comprehensively by measuring the S-I-S structure samples which have the various thickness of each layer. We have started from a search for an optimum thickness of the NbN layer of NbN-SiO$_2$-Nb multilayer sample. NbN-SiO$_2$-Nb multilayer samples with the various thickness of the NbN layer have been produced by ULVAC, Inc. [8,9]. In this report, measurement results of the bulk Nb sample and the NbN-SiO$_2$-Nb multilayer samples of the various thickness of the NbN layer and comparison with the theoretical prediction are discussed.

## MEASUREMENT SYSTEM

*Measurement Principle*

Let us consider the situation in which the solenoid coil is positioned above a Type-II superconducting sample, and an AC magnetic field ($H_{ap}$) is applied to the sample from the solenoid coil. A voltage of the solenoid coil is induced from both an AC current in the solenoid coil and a shielding current flows on the sample surface. If $H_{ap} < H_{c1}$, the sample maintains the Meissner state and response of the shielding current is linear. On the other hands, If $H_{ap} > H_{c1}$, the response of the shielding current becomes saturated and nonlinear, resulting in a magnetic flux to start penetrating the sample. This nonlinear response of the shielding current produces nonlinear voltage response in the solenoid coil, which generates a third harmonic voltage.

In this experiment, the sample is cooled down to the Meissner state with a zero magnetic field by an LHe stored in a bottom of a cryostat. Next, the sample is warmed up slowly (below 0.1 K/min) while applying a 1 kHz AC magnetic field from the solenoid coil to the sample. When a temperature of the sample exceeds a certain temperature, the magnetic flux starts to penetrate the sample. Then, the third-harmonic voltage (3 kHz) is induced in the solenoid coil. At this moment, we measure the effective $H_{c1}$ by detection of drastic change of the response of the third harmonic voltage. In addition, we plot a temperature dependence of the effective $H_{c1}$ by repeating the measurement of the effective $H_{c1}$ with various applied magnetic fields.

---

[†] hayatoi@post.kek.jp

## Sample and Coil Stages

A measurement setup consists of two copper stages, which have a diameter of 200 mm and a thickness of 5 mm, for the sample and solenoid coil, as shown in Fig. 1. The sample stage has two copper fins extended to the upper direction and each fin is equipped with a heater to increase the temperature of the sample. The coil stage also has two copper fins extended to the lower direction and whose bottom ends are immersed LHe to play a role of thermal anchor. The solenoid coil is positioned at the center of the coil stage and there are slits around the solenoid coil to prevent heat generation by eddy currents. Four Cernox sensors are used for temperature sensing. One of the temperature sensors is directly in contact with the rear surface of the sample through a hole in the center of the sample stage, and remaining temperature sensors monitor the temperature of any part of the measurement setup. The sample is put between the two stages. The gap distance of 0.05 mm between the sample surface and the solenoid coil is kept by 9 SiN balls embedded in the coil stage. The solenoid coil and SiN balls are fixed to the coil stage with epoxy adhesive.

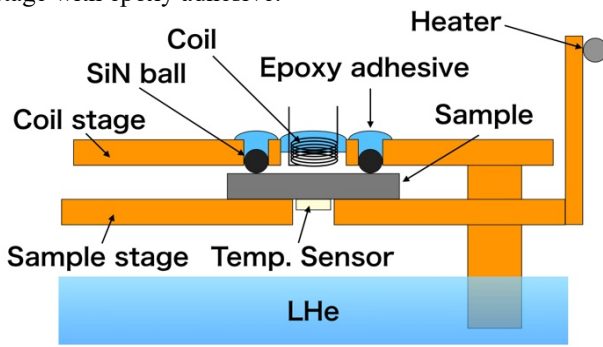

Figure 1: Cross-sectional schematic of the copper stage setup.

## Solenoid Coil

In order to measure the effective $H_{c1}$ in the low-temperature region, the solenoid coil which can apply the higher magnetic field is necessary. For example, the magnetic field of 150 mT is our target considering the enhancement of effective $H_{c1}$ of NbN-SiO$_2$-Nb multilayer samples. In addition, the solenoid coil needs to be sufficiently small relative to the sample size (50 mm × 50 mm). In order to determine the parameters of the solenoid coil that suffers these conditions, we simulated the magnetic field from the solenoid coil using Finite Element Method Magnetics (FEMM) [10].

As regards the simulation result, the peak value of the magnetic field on the sample surface was obtained as 111 mT with a current of 4.5 A reflecting the solenoid coil parameters which was used in this measurement (the inner diameter of the coil was 2 mm, outer diameter was 5 mm, length was 5 mm, and number of turns was 176). Since a limit of an amplifier which applies the AC current to the solenoid coil is 10 A, in principle, it is possible for the solenoid coil to produce the high magnetic field >150 mT. In addition, the magnetic field at the sample edge (25 mm) was 0.4 % of the peak value, thus we considered that the sample edge effect was negligible (see Fig. 2).

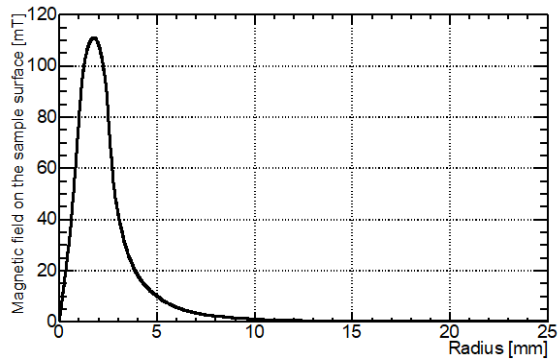

Figure 2: Magnetic field from solenoid coil on sample surface along radial direction.

## Circuit

Figure 3 shows a block diagram of the measurement circuit. A 1 kHz signal is first generated by a signal generator (S. G.) and fed into the amplifier via 1 kHz band-pass-filter (BPF). Next, the amplified 1 kHz signal is fed into the solenoid coil, then the 1 kHz AC magnetic field is applied to the sample. The amplified 1 kHz signal is detected by measuring the voltage across the solenoid coil and fed into a 3 kHz BPF. The detected 3 kHz signal is amplified with selectable gain values of 10, 100, and 1000 and acquired by means of an oscilloscope. Then, the fast Fourier transform (FFT) is performed to measure the third harmonic signal. The AC current flows in the solenoid coil which is used for magnetic field calibration is monitored by measuring the voltage across the 0.1-Ω resistor placed just after the amplifier. A 50% duty pulse operation (on-time 2 s, off-time 2 s) is performed to reduce heat generation of the solenoid coil.

## MASUREMENT RESULTS

### Result of Bulk Nb Sample

As magnetic field calibration and baseline test, the $H_{c1}$ measurement of the bulk Nb sample was performed. The standard surface treatment process for the SRF cavity was applied to the bulk Nb sample except for low-temperature baking. Figure 4 shows a typical measurement result of third harmonic response for bulk Nb sample. In order to determine the temperature at which the vortex starts to penetrate the sample, we applied the linear fitting to the third harmonic signal in the temperature region lower than the left onset (red line in Fig. 4). Next, the distribution of the difference between each point and the linear fitting function was examined. Finally, the first point which has the distance of 3σ from the linear fitting function was taken as the vortex penetration temperature. Consequently, in the case of Fig. 4, a vortex penetration temperature was determined as 5.3 K with a current of 4.4 A in the solenoid coil.

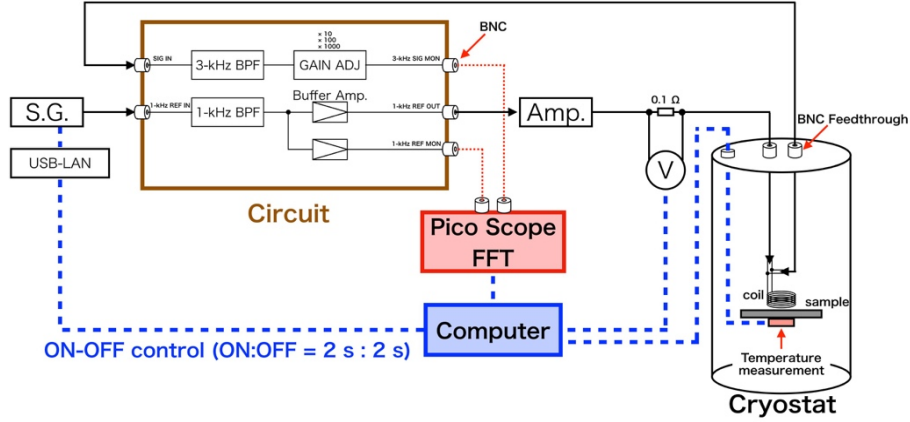

Figure 3: Block diagram of measurement circuit.

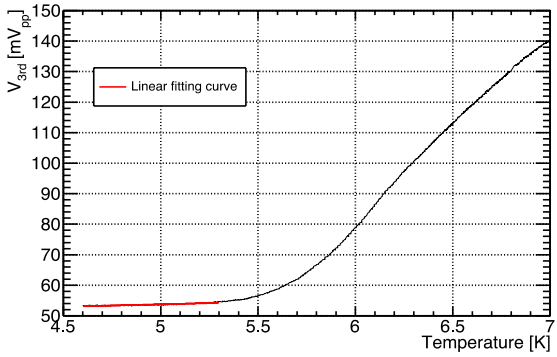

Figure 4: Third harmonic response vs. sample temperature of bulk Nb with the current of 4.4A in the solenoid coil. The vertical axis is the intensity of the third harmonic signal after FFT and the horizontal axis is the temperature of the bulk Nb sample.

The above analysis which determines the vortex penetration temperature was performed repeatedly at various AC currents. Figure 5 shows the measurement result of the relationship between the AC current and the vortex penetration temperature for the bulk Nb sample. Open circles represent the measurement points for the bulk Nb sample. The red curve represents the fitting curve using following function:

$$f(t) = a \times \left\{1 - \left(\frac{t}{b}\right)^2\right\} \quad (1)$$

where $a$ and $b$ are fitting parameters which correspond to the AC current at 0 K and the critical temperature $T_c$ of the bulk Nb sample, respectively. As regards the fitting result, each value was calculated to be $6.61 \pm 0.09$ A and $9.08 \pm 0.06$ K. Because the variation in the vortex penetration temperature for each measurement for the same magnetic field is at most 0.1 K, the temperature error for each open circle was uniformly determined as 0.1 K. The error of the AC current for each open circle was determined from the deviation of the current at each measurement point with in this temperature error.

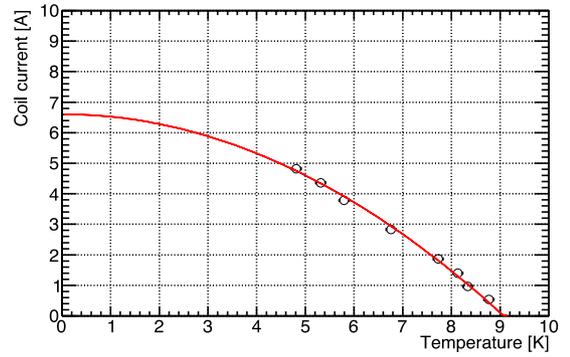

Figure 5: Relationship between the AC current and the vortex penetration temperature for the bulk Nb sample.

In order to obtain a calibration factor which converts the AC current to the magnetic field, we related the value of the AC current of $6.61 \pm 0.09$ A to the $H_{c1}(0)$ of Nb which is assumed to be 180 mT. Then, the temperature dependence of $H_{c1}$ for the bulk Nb sample was obtained by converting the AC current to the magnetic field with the use of the calibration factor (see open circles and black solid curve in Fig. 6).

*Result of NbN-SiO$_2$-Nb Multilayer Samples*

The same surface treatment as bulk Nb sample was performed to the Nb substrate of the NbN-SiO$_2$-Nb multilayer samples. The seven NbN-SiO$_2$-Nb multilayer samples with the SiO$_2$ layer of the 30 nm thickness and the various thickness of the NbN layer (50, 100, 150, 200, 250, 300, and 400 nm) were produced by ULVAC, Inc.. The six samples other than the 200 nm sample were produced at the same time, and only the 200 nm sample was produced about four months earlier than the other samples.

These samples were measured and the values of effective $H_{c1}$ of these samples were compared with the theoretical prediction. The analysis to determine the vortex penetration temperature was performed in the same way as the analysis for the bulk Nb sample and the temperature de-pendence of the effective $H_{c1}$ for each NbN-SiO$_2$-Nb multilayer sample was measured. Figure 6 shows the measurement result of the temperature dependence of the effective $H_{c1}$ for the NbN-SiO$_2$-Nb multilayer samples and comparison with the result of the bulk Nb sample. It can be seen that there is an enhancement of the effective $H_{c1}$ of the NbN-SiO$_2$-Nb multilayer sample as compared to the bulk Nb sample, and the value of the enhancement of the effective $H_{c1}$ is different for each NbN layer thickness.

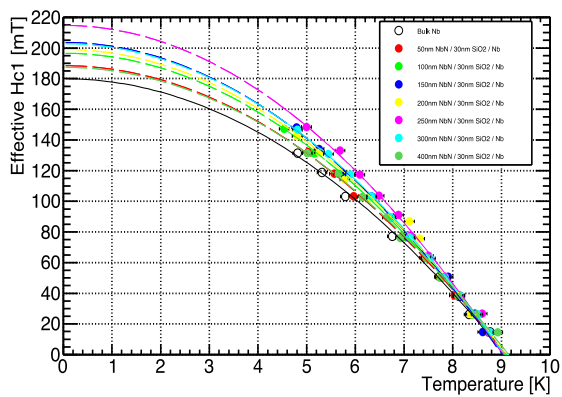

Figure 6: measurement result of the temperature dependence of the effective $H_{c1}$ for NbN-SiO$_2$-Nb multilayer and comparison with the result of bulk Nb sample.

A comparison of the value of the effective $H_{c1}$ of each NbN-SiO$_2$-Nb multilayer sample at 0 K with the theoretical prediction [11] is shown in Fig. 7. The open circle represents the measurement values of the effective $H_{c1}$ for the 200 nm sample during the development stage of the measurement setup. The measurement error for the 200 nm sample was 18 mT because the lower limit of the measurable temperature was at 8 K. The closed circles represent the measurement values of the effective $H_{c1}$ for each NbN-SiO$_2$-Nb multilayer sample after final tuning of the setup and each value was measured accurately within the error of at most 4 mT by improving the measurable temperature from 8 to 5 K. Note that the measurement values of the effective $H_{c1}$ for the 200 nm sample dropped from 226 ± 18 mT to 199 ± 4 mT. There are several possible reasons for this drop in the effective $H_{c1}$ of the 200 nm sample, i.e. the 200 nm sample was kept in the air for a longer period of time than other samples after production and was overused for the setup tuning. Four solid lines of different colors represent the theoretical curve with η = 1, 0.9, 0.8, and 0.7, respectively. η (< 0 < η ≤ 1) is a phenomenological suppression factor for NbN layer where η = 1 represents the ideal and smooth surface of the NbN layer, and if η becomes smaller more defects on the surface such as impurities and topographic defects etc. exist [12,13]. The peak value of the effective $H_{c1}$ decreases and the optimum thickness of the NbN layer shifts to the thinner direction as η decreases. The comparison between the measurement values and the theoretical curve show that optimum thickness, which maximizes the en-hancement of effective $H_{c1}$, exists for the NbN-SiO$_2$-Nb multilayer structure and our measurement results are in good agreement with the theoretical curve with η = 0.7 to 0.8 except for the result of the 200 nm sample after the final tuning of the setup. We also mention that the open circle for 200 nm sample is in good agreement with the theoretical curve. In the case of η = 0.7 to 0.8, the maximum improvement of 24 to 31 % for the NbN-SiO$_2$-Nb multilayer structure is expected compared with bulk Nb.

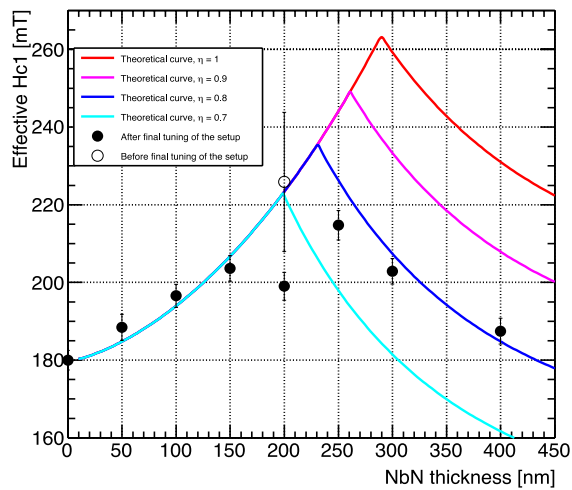

Figure 7: Effective $H_{c1}$ of NbN-SiO$_2$-Nb multilayer samples vs. thickness of NbN layer and comparison with theoretical prediction. The open circle represents the measurement values of the effective $H_{c1}$ for the 200 nm sample during the development stage of the measurement setup. The closed circles represent the measurement values of the effective $H_{c1}$ for each NbN-SiO$_2$-Nb multilayer sample after final tuning of the setup.

## CONCULUSION

The $H_{c1}$ measurement system using the third harmonic measurement method in KEK was constructed. The measurements of the bulk Nb sample and the NbN-SiO$_2$-Nb multilayer samples were performed successfully. As regards the measurement results, we found that the optimum thickness existed for the NbN-SiO$_2$-Nb multilayer structure. Further, our measurement results of NbN-SiO$_2$-Nb multilayer samples were in good agreement with the theoretical curve with η = 0.7 to 0.8 which corresponds to the maximum improvement of 24 to 31 % for the NbN-SiO$_2$-Nb multilayer structure compared with bulk Nb.

These results support that SRF cavity with the NbN-SiO$_2$-Nb multilayer structure has potential to achieve the higher accelerating gradient respect to conventional SRF cavity. Further, these results strongly suggest that the target parameters in the creation of the NbN-SiO$_2$-Nb multilayer structure on the inner surface of the SRF cavity were obtained and the possibility was opened that the production of the NbN-SiO$_2$-Nb multilayer cavity will be performed stably in mass production.

## ACKNOWLEDGEMENT

This work is supported by Japan Society for the Promotion of Science Grant-in-Aid for Young Scientist (A) No.17H04839.